# Understanding the Information needs of Social Scientists in Germany

**Dagmar Kern**
*GESIS – Leibniz-Institute for the Social Sciences, Cologne, Germany, dagmar.kern@gesis.org*

**Daniel Hienert**
*GESIS – Leibniz-Institute for the Social Sciences, Cologne, Germany, daniel.hienert@gesis.org*

**ABSTRACT**

The information needs of social science researchers are manifold and almost studied in every decade since the 1950s. With this paper, we contribute to this series and present the results of three studies. We asked 367 social science researchers in Germany for their information needs and identified needs in different categories: literature, research data, measurement instruments, support for data analysis, support for data collection, variables in research data, software support, networking/cooperation, and illustrative material. Thereby, the search for literature and research data is still the main information need with more than three-quarter of our participants expressing needs in these categories. With comprehensive lists of altogether 154 concrete information needs, even those that are only expressed by one participant, we contribute to the holistic understanding of the information needs of social science researchers of today.

## INTRODUCTION

The Social Sciences look at the social relationships between humans and their interactions within the society and incorporate disciplines such as Sociology, Demography, Education/Pedagogy, Psychology, Communication Studies, Economics, and Political Science (cp. Hogeweg-de Haart 1983). It is a broad research field with different actors, different interests, and different information needs. Already in 1979, Brittain (1979) discussed the specialties of the social sciences such as imprecise terminology, the absence of experiment replication, different schools of thought and a lack of consensus. Hogeweg-de Haart (1983) stated that the scope of the social sciences is broad, not well defined, and it varies a lot in different definitions and in different countries. Also, in each discipline potentially an unlimited number of themes could be addressed. Because of this and the interlinking of different disciplines, the variety of primary information types used is large. Additionally, secondary use of existing data plays an important role for social scientists and requires the search for archived research data for reuse. A couple of decades later, the scope of social science research has increased even further, and the introduction of digital media has opened new fields for activities, e.g., the upcoming field of computational social science (Gilbert 2010). Furthermore, there is a trend towards open science with open publications, open data and open methodology in the context of the FAIR-principles (findable, accessible, interoperable, re-usable, Wilkinson 2016). Applying the FAIR-principles enables a better way for replication and reuse of research outcomes but also cause new activities and new information needs on the side of the researchers.

In this paper, we present the results of three studies to capture the current information needs of social scientists in Germany. We performed a diary study in which 12 participants protocolled their information needs over a period of two weeks. In a second study, we presented 25 pre-formulated information needs in a questionnaire to 18 social science researchers and let them vote which information needs match their own. In the last study, 337 participants expressed their information needs in an online survey.

## RELATED WORK

There is a long tradition of studies to capture and understand the information use and needs as well as seeking behaviors of social scientists from the 1950s to nowadays. Folster (1995) summarizes the conclusions from three decades of research: (1) journals are the preferred information source for social scientists, (2) following citations is the preferred method, (3) informal channels play an important role, and (4) library services do not play a major role. Line (1971) reports on a study of information use and needs of social scientists in the UK of the 1970s. Data was gathered mainly by questionnaire (n=1,098), interviews (n=125) and daily observations. The results show the use of different information channels (mainly periodicals, then books, government publications, research reports, computer printouts, colleagues, etc.) but also different methods of locating this information (mainly bibliographies or references, then experts, abstracts and indexes, discussion, etc.).

Slater (1988) conducted thirty to forty interviews in the UK about the information needs of social scientists. Three interesting observations were that (1) a major library in the UK for the social sciences in general would be needed, that (2) there is a lack of a library dedicated just to social science methodology and that (3) online access to abstracting and indexing services was the way forward for the social sciences. Folster (1989) in 1987 surveyed social science researchers at University Wisconsin-Madison with 119 valid questionnaires. Similar to the study of Line (1971) they found out that journals are the most important source of information and tracking citations is the highest ranked method of seeking information. Computerized literature search was ranked lowest. De Tiratel (2000) examined the information use of social scientists in Argentina and found a similar information-seeking behavior as in prior studies of English-speaking countries such as the use of books and journals and the use of informal channels. Agrawal (1987) reports on similar information sources for social scientists in India. Shen (2007) conducted interviews with four professors of the social faculty of Wisconsin-Madison to identify information needs, sources and what causes problems in information search. She states that unlike research found before the 1990s, researchers make extensive use of electronic resources to find not only literature but also existing research datasets for secondary analysis. The major problems for researchers are that information is scattered in different places and that it is too much information.

Focusing on information seeking behavior, Ellis (1989) conducted interviews with academic social scientists and identified six broader categories of information seeking patterns: (1) starting (initial search for information), (2) chaining (following referential connections such as references), (3) browsing (e.g. by scanning of journals or table of contents), (4) differentiating (between sources to filter materials), (5) monitoring (the field of interest by following sources such as journals or conferences), and (6) extracting (relevant materials from sources). In a comparison study with physicists and chemists, Ellis et al. (1993) found two more categories: verifying (the accuracy of information) and ending (activities at the end of the project such as paper writing). Meho & Tibbo (2003) conducted e-mail interviews with sixty social science faculty members from 14 different countries to see if Ellis' model holds after the emergence of the World Wide Web and also on an international scale. They approved the model but could add four more categories: accessing (the material found with the methods above), networking (with colleagues to exchange information), verifying (the accuracy of the information), and information managing (by archiving and organizing information).

Identified already in earliest studies (e.g., Hogeweg-de Haart 1983), the re-use of research data for secondary analysis is a typical characteristic of social science research practice. Recently, with the upcoming of more and more data and online data platforms related work tries to understand the context and implications of data re-use. For the field of qualitative social science research Corti (2007) describes a number of factors that have the potential for the re-use of data sets: revisiting, reanalyzing and comparing with complementary data sources. For quantitative research this holds true, but here data mainly consists of survey data with attitudes, behaviors and factual information of a population group (Dulisch, 2015). Curty (2016) discusses benefits, risks, effort, social influence, facilitating conditions and reusability of data reuse based on interviews with 13 social scientists. Data re-use can avoid the time-consuming primary data collection and gathering process. Faniel et al. (2016) conducted a larger survey with 1,480 journal authors who have cited research data sets from the Inter-University Consortium for Political and Social Research (ICPSR). Data attributes such as completeness, accessibility, ease of operation, and credibility correlate positively with the re-users' satisfaction. The difference in the retrieval of literature and datasets is addressed in a work of Kern & Mathiak (2015). In a lab study and with telephone interviews, they found that the quantity and quality of the metadata are much more important in dataset retrieval. To identify if the data has the potential to answer their research question a lot of materials need to be scanned (codebook, questionnaire, raw data) which is much more time consuming than having a look in a paper to judge its relevance.

**DIARY STUDY**

With the goal to find real information needs of social scientists, we performed our first study - a diary study. The study took place in January and February 2016. We asked social scientists to protocol their information needs over a period of two weeks. We provided them a protocol that starts with an introduction, our contact information for questions, a couple of examples and a two column table. They were expected to write their current information need or the problem to be solved in the first column before they start to solve it and in the second column they ought to provide us information about the approach they used to meet their information need.

*Participants*

12 participants took part in the diary study (5 female, 7 male; mean age 37.5, SD=4.17). All of them are German social science researcher. 11 participants work at our research institute which focuses on empirical social science. One works as a professor at the University of Düsseldorf. Three of the participants have a doctoral degree, and the remaining 9 have a master degree or equivalent and are working on a Ph.D. project.

*Results*

Altogether, we collected 54 statements. On average, participants indicated 4.5 (SD=2.42) individual information needs. We categorized the statements (see table 1) and provide examples for each category and the approaches the subjects followed to fulfill the information need.

| Category | Number of statements (n=12) | Examples for information needs in this category | Approaches / Tools / Portals |
|---|---|---|---|
| Literature | 19 | From a specific author, on a specific topic, for a lecture, to get an overview, full texts | Web of Science, university library, ResearchGate, Google Scholar, Google, references of known/found papers, personal contacts, personal libraries, author's websites, library catalogs, digital libraries |
| Support for data analysis | 9 | On mappings, on metadata schemes, on network analysis, on classification of open answers | Looking for specific tools and specifications via Google, Google Scholar; personal contacts, YouTube, project websites |
| Research data | 4 | Data sets to a specific topic, that applied a specific test | Data catalogs, specialist literature, personal contacts |
| Illustrative material | 4 | Images, diagrams, videos | Web, Wikipedia, YouTube, Pixabay, |
| Definition /concepts | 3 | By a specific author, to a specific topic | Google (topic, author) |
| Networking /cooperation | 3 | Reviewers for a paper, journal, etc.; Who can give a talk on a specific topic | Sowiport, contacts |
| Support for data collection | 3 | Information about planning and designing a questionnaire, example questionnaires, recruiting of participants | Unipark, project websites, Google, mailing lists, Xing |
| Funding | 2 | Funding opportunities, What kind of research was already funded? | Institutional websites, Google |
| Strategy paper | 2 | On a specific topic | Specialist literature, personal contacts |
| Software support | 1 | On network analysis | YouTube, Google Books, personal contacts, support pages, user forum |
| Literature to a specific data set | 1 | Who used a specific data set? | Google |
| Reference lists | 1 | Geographic area data | Federal Statistical Office, Wikipedia |
| Measurement instruments | 1 | Scale to study personality traits | specialist literature, personal contacts |
| Training | 1 | Communication skills | Google, personal contacts |

**Table 1. Categories of information needs of social science researchers (n=12), number of statements in each category, examples and approaches to fulfil the information need.**

## QUESTIONNAIRE ON INFORMATION NEEDS

In a second step, we generated a list of 25 more general information needs based on the results of the diary study and on experiences of colleagues at our institute who provide specific services for social science researchers, e.g., they support them by finding appropriate data, designing questionnaires or analyzing survey research data. We presented this list in paper form to 18 different social science researchers in the context of a user study in March 2016. Table 2 shows the whole list in the column "I'm looking for …". The participants had to check which information needs correspond to their own.

### Participants

18 German social scientists (7 female, 11 male, mean age 33.35, SD=10.04) took part in this study. Participants were recruited through personal E-mail invitations sent to the employees of the School of Social Science at the University of Mannheim and to the employees of the Institute of Social Science at the University of Düsseldorf. Two participants work as professors, four are postdoctoral researchers, and the remaining 12 held a master degree and work as research associates. 14 work with quantitative data, three with quantitative and qualitative data and one with neither of them. Eight of them narrow down their field of expertise to sociology, four perform research in the field of political science, two in psychology, two in survey methodology and two in communication science.

*Results*

In table 2 the comprehensive list of information needs is shown with the number of matches to the personal information need. For better comparison of the studies, we assigned each information need to categories also used in table 1. These categories were not presented to the participants in the study.

| I'm looking for … | Number (n=18) | Category |
|---|---|---|
| … full texts, which I can download directly. | 15 | Literature |
| … current literature by a certain author. | 14 | Literature |
| … literature in general. | 13 | Literature |
| … references of a paper. | 13 | Literature |
| … literature that gives me an overview of the topic. | 13 | Literature |
| … questionnaires on a specific topic. | 12 | Support for data collection |
| … variables collected in various studies to compare them. | 12 | Variables in research data |
| … international research data. | 11 | Research data |
| … literature that reports on analyses of specific research data. | 11 | Literature |
| … access to certain research data. | 11 | Research data |
| … information about which variables are included in the research data. | 11 | Variables in research data |
| … information on items and scales contained in research data. | 11 | Measurement instruments |
| … social science items and scales to create a questionnaire. | 11 | Measurement instruments |
| … information about training / seminars / summer schools. | 11 | Training |
| … national research data. | 10 | Research data |
| … study description to specific research data. | 10 | Research data |
| … research data mentioned in a paper. | 9 | Research data |
| … research projects. | 9 | Networking / Cooperation |
| … links between research data. | 7 | Research data |
| … questions for a study that I would like to carry out. | 7 | Support for data collection |
| … a definition of a term by a certain author. | 6 | Definition |
| … information on data analysis of microdata. | 6 | Support for data analysis |
| … information about questionnaire construction and development of survey tools. | 6 | Support for data collection |
| … material to create a lecture. | 5 | Illustrative material |
| … information on methodological aspects of organizing and conducting surveys. | 3 | Support for data collection |

**Table 2. Information needs presented to 18 social science researchers and the number of people who indicated that the information need corresponds to their own. The last column shows the category the information need is afterward assigned to.**

**ONLINE-QUESTIONNAIRE**

In the third and last study, we created an online questionnaire in German to collect further insights from a larger group of social science researchers. We asked two questions only: (1) "In the following list, you find information needs of social sciences researchers. Please check the information needs that correspond to your own information needs.". The list options are shown in table 3 under "I'm looking for…". (2) "If you're thinking about the last week, what information were you looking for in the context of your scientific work? You are welcome to be concrete at this point, e.g. I was looking for instructions on how to categorize open replies". The second question was an open question and the first question primarily served to convey an idea of what kind of information needs is being addressed. The questionnaire was online from 25[th] of October 2017 until 31[st] of December 2017. The goal of this questionnaire was to collect more specific and individual information needs to get a more holistic view on social science information needs.

## Participants

We sent invitations through a mailing list containing social science researchers in Germany, who agreed to be contacted again in a prior study. The original mailing list was created manually by our institute. We went through all web presences of Social Science institutes at German Universities and collected contact details from employees, who held at least a Master degree. Thus, we were able to send personalized email-invitation to 1,107 researchers. 223 of them answered the online questionnaire completely (response rate of 20%). Furthermore, we published an invitation to the study as a news entry on our homepage and distributed it through the social media accounts (Facebook and Twitter) assigned to our institute. Thus, we received further 114 filled out questionnaires. Altogether, we got 337 exploitable questionnaires.

## Results

Table 3 shows the list of given information needs and the frequency of how often participants indicated this information need as their own. In the last column again the assigned category is shown, which was not presented to the participants in the study. As the information needs of the two participant groups do not differ significantly, they are presented together.

| I'm looking for | Number (n=337) | Category |
|---|---|---|
| ... literature on a specific topic. | 293 | Literature |
| … for full texts, which I can download directly. | 287 | Literature |
| ... national and/or international research data. | 216 | Research data |
| ... social science items and scales to create a questionnaire. | 171 | Measurement instruments |
| ... information about which variables are contained in certain research data. | 163 | Variables in research data |
| … for clues as to whether certain research data are suitable for answering my research question. | 159 | Research data |
| … variables collected in various studies to compare them. | 154 | Variables in research data |
| ... research data mentioned in a paper. | 134 | Research data |
| ... for cooperation partners for a research project. | 62 | Networking / Cooperation |

**Table 3. The first question in our online questionnaire on information needs and the number of participants who indicated that the provided information need corresponds to their own.**

187 participants also answered the second question. Altogether, we got 331 individual information needs. We were able to assign 305 of them to the already introduced categories: literature (see table 4), research data (see table 5), measurement instruments (see table 6), support for data analysis (see table 7), support for data collection (see table 8), variables in research data (see table 9), software support (see table 10), networking / cooperation (see table 11) and illustrative material (see table 12). 26 information needs could not be assigned to any category as their meaning was not obvious. The *n* in the column "number" refers to the number of persons who have made one or more statements in this category.

| Literature | Number (n=98) |
|---|---|
| Literature in general | 81 |
| Full text | 22 |
| Current literature | 4 |
| Literature of a specific author | 2 |
| Literature on fundamental work | 2 |
| Literature on related topics | 2 |
| Collection on sources | 2 |
| Cited literature | 2 |
| E-books, English literature, scientific journals, grey literature, journal articles, secondary literature, full text from journals/books, literature from related disciplines, literature on method comparison, literature to research data, survey articles, research reports, book reviews, methodical literature | Mentioned once in each case |

**Table 4. Information needs in the category literature.**

| Research data | Number (n=61) |
|---|---|
| Research data | 36 |
| Research data for comparison | 8 |
| Survey data | 5 |
| Historical data / time series | 3 |
| Research data mentioned in a paper | 2 |
| Statistics | 2 |
| Research data from a specific country | 2 |
| Access to research data | 2 |
| Data on population, research data from a specific country, geodata, microdata, qualitative data, linked data | Mentioned once in each case |

**Table 5. Information needs in the category research data.**

| Measurement instruments | Number (n=26) |
|---|---|
| Scales to a specific question | 6 |
| Items to a specific question | 5 |
| Information about specific scales | 5 |
| Items and variables on which my projects can be based | 3 |
| Items and scales from validated questionnaires | 3 |
| I searched for questionnaires containing items for specific concepts | 1 |
| Translation of scales | 1 |
| Replicability/copyright of a scale | 1 |
| Operationalization of scales and items | 1 |

Table 6. Information needs in the category measurement instruments.

| Support for data analysis | Number (n=23) |
|---|---|
| Information on a specific analysis method | 7 |
| Practical instructions for qualitative content analysis | 3 |
| Mathematical assumptions of maximum likelihood method | 1 |
| Literature on the interpretation of multinomial logistic regression and fixed effects models | 1 |
| Methodology literature to adjust my data analysis to the latest development | 1 |
| Instructions on statistical methods for answering my research question | 1 |
| Texts on the application of machine learning techniques in the social sciences. | 1 |
| Calculations of certain variables / syntax files | 1 |
| Application examples for a statistical method | 1 |
| Methodological articles on community analyses of hierarchical and mixed models | 1 |
| Documents for item analysis with R | 1 |
| Application possibilities of network analyses | 1 |
| Stata-DoFiles and keys for converting between different professional codes | 1 |
| Macro-indicators for international comparative studies | 1 |
| Instructions for comparative methodology | 1 |

Table 7. Information needs in the category support for data analysis (table continued from previous page).

| Support for Data collection | Number (n=14) |
|---|---|
| Information on survey methods | 5 |
| Instructions on how to create a survey | 4 |
| Formulation for item batteries | 3 |
| Recruiting of participants | 2 |
| How to create an online questionnaire with Unipark | 1 |
| Drawing of samples | 1 |
| Collection methods for qualitative surveys | 1 |
| Information on transnational standards for recording educational attainment | 1 |

Table 8. Information needs in the category support for data collection.

| Variables in research data | Number (n=13) |
|---|---|
| Specific variables collected in ALLBUS/ISSP | 2 |
| Variables including item formulation in codebooks | 1 |
| Certain variables in a known dataset (wanted to know if a particular question was asked last year) | 1 |
| Literature on a specific dependent variable (Which articles have examined this variable) | 1 |
| Variables in different datasets | 1 |
| Availability of certain variables in international population surveys | 1 |
| How have other studies used several variables on a topic | 1 |
| Variables contained in a research data set | 1 |
| Information on the frequency of collection of certain Eurobarometer questions | 1 |
| I checked whether certain items relevant to my question are available in large datasets | 1 |
| Variables in data sets | 1 |
| Descriptions of the variables contained in research data | 1 |
| Specific variables to answer a research question | 1 |

**Table 9. Information needs in the category variables in research data.**

| Software support | Number (n=12) |
|---|---|
| Stata Tutorials | 4 |
| Software for processing large data sets | 1 |
| Software for testing hypotheses on longitudinal data | 1 |
| Tool for visualization of statistical distributions | 1 |
| Software for the analysis of social networks, through website analysis | 1 |
| Tool for merging macro dataset (country level) with a micro dataset (EQLS 2003-2011) | 1 |
| I was looking for pros and cons of the Python programming language for statistical analysis | 1 |
| Tutorial for statistical programs | 1 |
| Help for dealing with MAXQDA | 1 |
| R-Base | 1 |

**Table 10. Information needs in the category software support.**

| Networking / Cooperation | Number (n=6) |
|---|---|
| Project partners | 2 |
| Scientific project to a specific topic | 1 |
| Overview of research projects on specific research questions. | 1 |
| Who is currently working with the data (in a project)? | 1 |
| Possible referent for a specific topic | 1 |
| Call for papers (conferences) | 1 |
| Online forums on specific scientific methods | 1 |
| I was looking for experts on certain aspects of the content who were willing to cooperate. | 1 |

**Table 11. Information needs in the category networking / cooperation.**

| Illustrative material | Number (n=3) |
|---|---|
| Video, images | 1 |
| Definitions | 1 |
| Lecture material | 1 |

**Table 12. Information needs in the category illustrative material.**

## DISCUSSION

By conducting three different studies (diary, questionnaire, online survey), we were able to collect 154 different information needs from 367 social science researchers in Germany. Our focus was on collecting concrete information needs that arise in the daily working life of social science researchers. The advantage of the mixed-mode method is that new real information needs can be identified on a high scale. Our approach is a bit different to prior studies which often ask researchers which predefined information types they use (journals, books, etc.) and how they are accessed (online, offline, etc.). Although the method implicates a diversity of answers, we were able to cluster most of the results into nine different groups: (1) literature, (2) research data, (3) measurement instruments, (4) support for data analysis, (5) support for data collection, (6) variables in research data, (7) software support, (8) networking/cooperation, and (9) illustrative material.

(1) Searching for literature is still the unchallenged main information need of researchers in the field of social science. In all studies, more than 80% of the participants stated at least one information need in this category: in the diary study 83%, in the offline questionnaire 88%, in the online questionnaire first question (closed) 90% and the second question (open) 52%. In early work (Foster 1995, Line 1971, De Tiratel 2000), researchers mentioned journals as their most important source of information. The results of our studies show that the sources itself seem to play a minor role by now whereas accessibility to full texts is becoming more and more important.

(2) The re-use of existing research data is a specialty of the social sciences which has been identified already in early studies (e.g. Hogeweg-de Haart 1983). With more and more data available online this information need gets more prominent. In the period of our diary study, only two of the 12 people searched for research data and also in the online questionnaire when we asked for the information need of last week only 32% stated "searching for research data" as their current information need. However, asking with a closed question about their general information needs as we did it in the questionnaire study and with the closed question in the online study the percentage of persons who are looking for research data is at 88% (questionnaire study) respectively 75% (online study).

Regarding the approaches social science researchers fulfill their information need; we found out, that all of the 12 participants in our diary study used the web at least as a starting point for most of their information tasks. As the most often used search engine "Google" was mentioned by ten subjects. Five participants used Google Scholar at the time of the study. Seven participants used discipline-specific databases or digital libraries such as Web of Science or Scopus. Beside of this, the following digital sources are mentioned: institutes' or authors' websites, ResearchGate, YouTube, Wikipedia, Amazon, and Pixabay. Only 3 mentioned non-digital sources like personal contacts (3), (University) libraries (2) and the personal book collection.

Although the way to get literature and research data has changed in the last decades, the interest in these categories is still and probably will remain high in the future. Especially, our participants in the questionnaire study expressed the desire for open access, 15 of 18 participants already look for full text that they can download directly. Also the need for linked information, e.g. research data mentioned in a paper or literature that reports on analyses of specific research data are information needs by more than half of the participants.

(3) Measurement instruments in the field of social science are used to collect sociological attributes in a standardized way. The instruments are items and scales that can be applied in surveys and enable the comparability of research data. In our diary study, one participant looked, for example, for a scale to study personality traits. In the online study, we addressed this category with two information needs which were checked by 11 participants each (61%). In the online study, also 50% checked the corresponding information need in the closed question, and 26 participants provided insight on their concrete information need in this category. This category and the following ones have not been identified as explicit information needs of social scientists in prior studies.

(4) The support for data analysis was the second most often mentioned information need in our diary study. The participants looked for mappings, metadata schemes, network analysis or classification of open answers. In our questionnaire study, we only asked for the information need of analyzing microdata which was checked by 6 participants. In the online study, 23 participants (12%) gave us detailed information on their information need in this category. The information needs in this category are manifold and in some cases very specific, but they show the general desire for better support of this information need.

(5) Three participants in our diary study expressed the need for support in the data collection process. They were looking for help regarding the planning of the study, designing the questionnaire and recruiting participants. In the questionnaire study, seven participants checked information needs in this category (39%). In the online study, this information need was not addressed in the close question, but 14 participants (7%) formulated their information needs in this category in the open question.

(6) The information need "variables in research data" was not mentioned in the diary study but added to the list of information needs presented to the participants in the questionnaire study because of the upcoming trend of variable search (such as provided by ICPSR[1]). Answers given in the questionnaire study as well as in the online study shows that about 50% of the researchers have this information need and 13 participants in the online study provide detailed information on their information need in the category variables.

(7) In the diary study, one subject reported the need for software support, in the online survey 12 (6%). Researchers in the quantitative social sciences, but also in the qualitative science use software to analyze their data. This process obviously needs support. (8) Networking/cooperation is a basic need of social scientists with three reported needs in the diary study, nine in the questionnaire, and 62 in the online closed question (18%) and 6 in the online open question (3%). (9) Illustrative material was reported to be needed in the diary study by four persons and online in the open question by 3 (2%).

Studies on information needs (in the social sciences) were conducted all time with the purpose to improve access to information and to adapt it to the current needs. Literature could be a long time only accessed through the library, the upcoming of the web and the trend of open access has adopted this behavior. Research data is more and more requested, and the rising of web portals for this purpose shows the trend. Very recently, also specialized portals came up which allow the search for variables in research data which also showed to be an information need in our study. Still underestimated is the search for reusable measurement instruments. Also, this trend shows up in the open science movement with the term open methodology, comprehensive search portals which supports this need are missing. Support for data analysis, data collection and software support are often very individual and at least at our institute addressed with personal consulting and training. However, the barrier for this is high, and users look for quick help on the Web. These categories hold larger information needs that are not yet addressed by specialized portals.

## PRACTICAL IMPLICATION

The presented studies were conducted as parts of a user-centered design process[2] which was applied to develop an integrated search system (https://search.gesis.org). This search system provides access to different information in the field of Social Science trough one single search bar. The search results are divided into different categories according to the findings of the conducted user studies, namely: *research data* (includes German studies retrieved from the GESIS data catalog as well as metadata of international studies), *publication* (includes open access publications and literature linked to research data), *questions & variables* (from research data archived in the GESIS data catalog), *instruments & tools* (includes measurement instruments, support for data collection and data analysis, and software support). Furthermore, the search system provides access to *GESIS Websites* on which users can find further support for conducting and analyzing survey data as well as to publications that are available in our *library*. The least demanded categories networking/cooperation and illustrative material are not included. One main benefit of the integrated search system is that information items are linked to each other so that users can see, for example, which publications contain data citations to research data. As we have seen, the information needs of social science are manifold and sometimes very individual and specific - the new integrated search system tries to support the most frequently mentioned.

## ACKNOWLEDGMENTS

We would like to thank all who have participated in our studies.

## REFERENCES

Agrawal, S. P., & Lal, M. (1987). Information needs of social scientists. International Library Review, 19(3), 287–299. https://doi.org/10.1016/0020-7837(87)90039-2

Brittain, J. (1979). Information Services and the Structure of Knowledge in the Social Sciences. International Social Science Journal, 31(4), 711–28.

---

[1] https://www.icpsr.umich.edu/icpsrweb/ICPSR/ssvd/index.jsp
[2] according to ISO 9241-201:2010 https://www.iso.org/standard/52075.html

Corti, L. (2007). Re-using archived qualitative data – where, how, why? Archival Science, 7(1), 37–54. https://doi.org/10.1007/s10502-006-9038-y

Curty, R. G. (2016). Factors Influencing Research Data Reuse in the Social Sciences: An Exploratory Study. IJDC, 11, 96–117.

de Tiratel, S. R. (2000). Accessing information use by humanists and social scientists: a study at the Universidad de Buenos Aires, Argentina. The Journal of Academic Librarianship, 26(5), 346–354.

Dulisch, N., Kempf, A. O., & Schaer, P. (2015). Query Expansion for Survey Question Retrieval in the Social Sciences. In S. Kapidakis, C. Mazurek, & M. Werla (Eds.), Research and Advanced Technology for Digital Libraries (pp. 28–39). Cham: Springer International Publishing.

Ellis, D., Cox, D., & Hall, K. (1993). A comparison of the information seeking patterns of researchers in the physical and social sciences. Journal of Documentation, 49, 356–356.

Ellis, David. (1989). A behavioural approach to information retrieval system design. Journal of Documentation, 45(3), 171–212.

Faniel, I. M., Kriesberg, A., & Yakel, E. (2016). Social scientists' satisfaction with data reuse. Journal of the Association for Information Science and Technology, 67(6), 1404–1416.

Folster, M. B. (1989). A study of the use of information sources by social science researchers. Journal of Academic Librarianship, 15(1), 7–11.

Folster, M. B. (1995). Information seeking patterns: social sciences. The Reference Librarian, 23(49–50), 83–93.

Hogeweg-de Haart, H. P. (1983). Characteristics of social science information: A selective review of the literature. Part I. Social Science Information Studies, 3(3), 147–164. http://dx.doi.org/10.1016/0143-6236(83)90021-2

Kern, D., & Mathiak, B. (2015). Are There Any Differences in Data Set Retrieval Compared to Well-Known Literature Retrieval? In S. Kapidakis, C. Mazurek, & M. Werla (Eds.), Research and Advanced Technology for Digital Libraries (pp. 197–208). Cham: Springer International Publishing.

Line, M. B. (1971). The information uses and needs of social scientists: An overview of INFROSS. In Aslib proceedings (Vol. 23, pp. 412–434). MCB UP Ltd.

Meho, L. I., & Tibbo, H. R. (2003). Modeling the Information-seeking Behavior of Social Scientists: Ellis's Study Revisited. J. Am. Soc. Inf. Sci. Technol., 54(6), 570–587. https://doi.org/10.1002/asi.10244

Gilbert, G. Nigel, ed. *Computational social science*. Vol. 21. Sage, 2010.

Shen, Y. (2007). Information seeking in academic research: A study of the sociology faculty at the University of Wisconsin-Madison. Information Technology and Libraries, 26(1), 4.

Slater, M. (1988). Social Scientists' Information Needs in the 1980s. Journal of Documentation, 44(3), 226–237.

Wilkinson, M.D. et al.: The FAIR Guiding Principles for scientific data management and stewardship. Sci. Data. 3, 160018 (2016).